%% file: CGlightning.tex
\documentclass[10pt,a4paper]{article}

\usepackage{pgfplots}
\pgfplotsset{compat=newest}

\usepackage{upgreek}
\usepackage{enumerate}
\usepackage{tikz}
\usepackage[small,margin=12mm]{caption}

\newcommand{\Xp}{\ensuremath{\mathrm{p}}}
\newcommand{\Xpp}{\ensuremath{\mathrm{p}^+}}
\newcommand{\Xpi}{\ensuremath{\uppi}}
\newcommand{\Xpip}{\ensuremath{\uppi^{+}}}
\newcommand{\Xpim}{\ensuremath{\uppi^{-}}}
\newcommand{\Xpipm}{\ensuremath{\uppi^{+/-}}}

\newcommand{\Xmu}{\ensuremath{\upmu}}

\newcommand{\Xmupm}{\ensuremath{\upmu^{+/-}}}
\newcommand{\XK}{\ensuremath{\mathrm{K}}}

\newcommand{\XKpm}{\ensuremath{\mathrm{K}^{+/-}}}
\newcommand{\Xe}{\ensuremath{\mathrm{e}}}

\newcommand{\Xv}{\ensuremath{\upnu}}

\newcommand{\dydx}[2]{\ensuremath{\frac{\mathrm{d}#1}{\mathrm{d}#2}}}

\newcommand{\midx}[2]{\ensuremath{#1_\mathrm{#2}}}
\newcommand{\Wk}{\midx{W}{k}}
\newcommand{\Wc}{\ensuremath{W_\mathrm{c}}}

\newcommand{\average}[1]{\ensuremath{\langle#1\rangle}}

\newcommand{\collision}[2]{\ensuremath{#1+\mathrm{X}\to\mathrm{X}^{*}+#2}}
\newcommand{\decay}[2]{\ensuremath{#1\to#2}}
\newcommand{\nx}[1]{\ensuremath{n_{#1}\cdot#1}}

\title{About Geometry and Initial Phase of\\ Cloud-to-Ground
Lightning}
\author{Ale\v{s}~Berkopec\footnote{\footnotesize{University of Ljubljana, Faculty of Electrical Engineering, Tr\v{z}a\v{s}ka 25, 1000 Ljubljana, Slovenia, {\slshape ales.berkopec@fe.uni-lj.si}}}}
\date{\today}

\begin{document}
\maketitle
\begin{abstract}
Cloud-to-ground lightning is the most common among
atmospheric discharges. Since electric fields in the
vicinity of a thunder-cloud do not exceed 250~kV/m the
physical process that triggers the lightning remains
unexplained.~\cite{FreeCharge,marshall1991electric,winn1974measurements}
Recent measurements established a weak correlation between
solar wind and incidence of lightning.~\cite{ScottSolar}
Here we show, that if an ionized path created by cosmic rays
provides a trigger, the distribution of lengths between two
successive forking points in a lightning channel
(internodes) closely resembles the exponential distribution
with average length between 415~m and 510~m. The results, if
confirmed, imply that a thunder-cell may be an additional 
source of fast elementary particles that initiate lightning
process.
\end{abstract}

\section{Reaction types and lightning topology}

A charged particle creates an ionized path when
passing through the atmosphere. Any collisions of the
projectile with nuclei in the atmosphere may produce
additional projectiles creating a fork in the path.
The result of this process is a tree structure of 
ionized paths, whose geometry matches the geometry 
of the stepped leader and that of subsequent lightning. 

The charged projectiles in cosmic showers are protons
\Xpp, pions \Xpipm, kaons \XKpm, and muons \Xmupm.~\cite{pdg2012booklet}
The roles of electrons and positrons are here neglected
due to at least four orders of magnitude lower ionization
rates.

We group the reactions of projectiles with nuclei
$\mathrm{X}$ in air according to their correspondence to
parts of a lightning channel. Notation \{a,b,c\} here
refers to "a, b, or c", and symbols for particles denote their
charged variants, for example \Xpi~denotes either \Xpip~or
\Xpim.

\newcommand{\mpta}{0.15\textwidth}
\newcommand{\mptb}{0.75\textwidth}
\begin{enumerate}[a)]
\item stem, end of a channel in mid-air, corresponds to 
	one of the following cases:\\
\vskip-10pt
\begin{tabular}{ll}
\begin{minipage}{\mpta} \input{f0.tikz} \end{minipage}
\begin{minipage}{\mptb}
	\begin{itemize}
	\setlength{\itemsep}{-2pt}
	\item particle \{\Xp,\Xpi,\XK\,\Xmu\} coming to rest, or
	\item capture of the charged projectile 
		\collision{\{\Xp,\Xpi,\XK\}}{\mathrm{neutrals}}, or
	\item decay \decay{\Xmu}{\Xe+\Xv+\bar{\Xv}}
	\end{itemize}
\end{minipage}
\end{tabular}
\item non-forked part corresponds to one of the
	following cases:\\
\vskip-10pt
\begin{tabular}{ll}
\begin{minipage}{\mpta} \input{f1.tikz} \end{minipage}
\begin{minipage}{\mptb}
	\begin{itemize}
	\setlength{\itemsep}{-2pt}
	\item passage of \{\Xp,\Xpi,\XK\,\Xmu\} without decay or collision, or
	\item collision of swap type
	\collision{\{\Xp,\Xpi,\XK\}}{\{\Xp,\Xpi,\XK,\Xmu\}+\mathrm{neutrals}}, or
	\item decays\\
		\phantom{xx}\decay{\Xpi}{\Xmu+\Xv}\\
		\phantom{xx}\decay{\XK}{\Xpi+\mathrm{neutrals}}
	\end{itemize}
\end{minipage}
\end{tabular}
\item forking with $N=2$ prongs corresponds to\\
\vskip-10pt
\begin{tabular}{ll}
\begin{minipage}{\mpta} \input{f2.tikz} \end{minipage}
\begin{minipage}{\mptb}
	collision
	\collision{\{\Xp,\Xpi,\XK\}}
	{\nx{\Xp}+\nx{\Xpi}+\nx{\XK}+\nx{\Xmu}+\mathrm{neutrals}},
	where $n_{\Xp}+n_{\Xpi}+n_{\XK}+n_{\Xmu}=2$
\end{minipage}
\end{tabular}
\item forking with $N=3$ prongs corresponds to either\\
\vskip-10pt
\begin{tabular}{ll}
\begin{minipage}{\mpta} \input{f3.tikz} \end{minipage}
\begin{minipage}{\mptb}
	\begin{itemize}
	\setlength{\itemsep}{-2pt}
	\item collision
		\collision{\{\Xp,\Xpi,\XK\}}
		{\nx{\Xp}+\nx{\Xpi}+\nx{\XK}+\nx{\Xmu}+\mathrm{neutrals}}, 
		where $n_{\Xp}+n_{\Xpi}+n_{\XK}+n_{\Xmu}=3$, or
	\item decay \decay{\XK}{\Xpi+\Xpi+\Xpi}~(charged pions)
	\end{itemize}
\end{minipage}
\end{tabular}
\item forking with $N\ge4$ prongs corresponds to\\
\vskip-10pt
\begin{tabular}{ll}
\begin{minipage}{\mpta} \input{f4.tikz} \end{minipage}
\begin{minipage}{\mptb}
	collision
	\collision{\{\Xp,\Xpi,\XK\}}
	{\nx{\Xp}+\nx{\Xpi}+\nx{\XK}+\nx{\Xmu}+\mathrm{neutrals}},
	where $n_{\Xp}+n_{\Xpi}+n_{\XK}+n_{\Xmu}\ge4$
\end{minipage}
\end{tabular}
\end{enumerate}
A fork can be produced only by collision of a charged hadron
with a nucleus or, in case of kaon ($N=3$ case), a decay
into three charged pions.
An internode part of the channel between two subsequent
forking points can only be created by one of the charged hadrons
\Xp, \Xpi, or \XK.

\section{Range and ionization rate}
\newcommand{\dd}[1]{\ensuremath{\mathrm{d}#1}}
\newcommand{\bpar}{\ensuremath{\kappa}}
\newcommand{\bpaq}{\ensuremath{\alpha}}

The charged projectiles interact with electrons in the surrounding
media loosing their kinetic energy according to Bethe
formula.~\cite{bethe1934stopping} 
In non-dimensional form with energy normalized as
$y=\gamma-1=\Wk/(m_0c_0^2)$ and distance $l$ normalized as
$x=l/\lambda$ the relativistic variant of the formula reads: 
\begin{equation}
\dydx{y}{x}=-\frac{\lambda\,\kappa}{m_0\,c_0^2}\cdot
\bigg[\frac{(y+1)^2}{y(y+2)}\cdot\ln\frac{\bpaq y(y+2)}{
\sqrt{1+2\mu(y+1)}}-1\bigg]
\label{eq:bethe}
\end{equation}
\begin{figure}[htb]
\centerline{\input{range.tikz}\hskip0mm\input{qe.tikz}}
\caption{Range $L$ for particles as function of initial energy 
	$W_1$ (left) and electric charge density $q_\mathrm{e}$
	on the trajectory as a function of the remaining
	path length. Both graphs are valid in the lower Earth
	atmosphere.} \label{fig:rangeqe}
\end{figure}
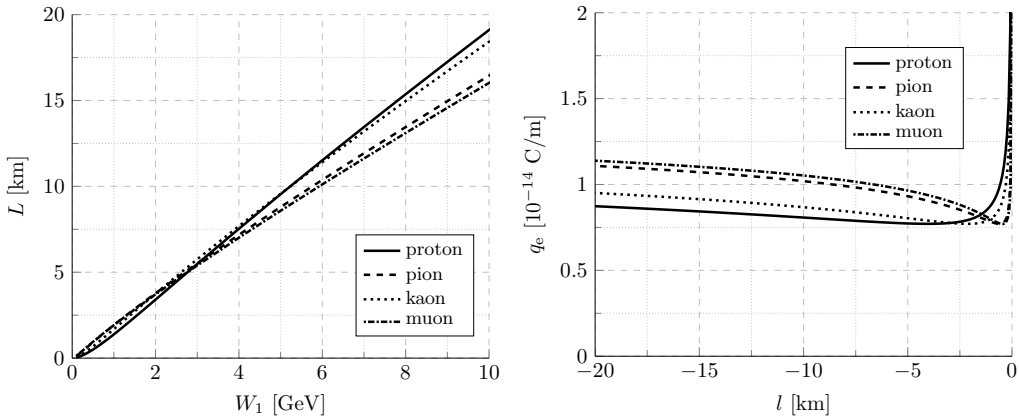
where $m_0$ is the rest mass of the projectile, $\mu$ is
ratio of electron mass to projectile mass, and $\lambda$ is 
its mean free path.
The mean free paths for the projectiles modelled as electrically 
neutral rigid balls with dimensions corresponding to their charge 
radii give
$\lambda_{\Xp}=417$~m, 
$\lambda_{\Xpi}=444$~m, and
$\lambda_{\XK}=507$~m.
The parameters \bpar~and \bpaq~are
\begin{eqnarray}
\nonumber
\bpar&=&\frac{\mathrm{e}_0^4\,n_{\mathrm{e}}}{2\uppi\,m_\mathrm{e}\,c_0^2
\,\varepsilon_0^2}
\approx4.77\cdot10^{-13}~\frac{\mathrm{J}}{\mathrm{m}}\approx
2.98~\frac{\mathrm{MeV}}{\mathrm{m}}\\
\bpaq&=&\frac{2\,m_\mathrm{e}\,c_0^2}{\average{I}}
\approx1.28\cdot10^4
\end{eqnarray}
with the following values for the lower Earth atmosphere: 
$\mathrm{e}_0=1.6\cdot10^{-19}$~A\,s,
$n_\mathrm{e}=4.59\cdot10^{26}$~m$^{-3}$,
$c_0=3\cdot10^{8}$~m/s,
$\varepsilon_0=8.854\cdot10^{-12}$~A\,s/(V\,m), and
$\average{I}=72.663$~eV.

For any type of the projectile the charge deposited on the
trajectory at energies above 0.5~GeV is around
$10^{-14}$~C/m (see Fig.~\ref{fig:rangeqe}, right). The distance
between a thunder-cell and ground is above 2~km. To
reach this distance, a projectile requires initial kinetic
energy of at least 2~GeV (see Fig.~\ref{fig:rangeqe}, left).

\section{Average internode length}
\newcommand{\Pc}{\midx{P}{col}}
\newcommand{\Pd}{\midx{P}{dec}}


The probability that a neutral particle collides with a nucleus
inside a cube of side length $\Delta{l}$ is approximately
$q_0=\Delta{l}/\lambda$ for $\Delta{l}\ll\lambda$. 
For charged particles this value changes
to $q/q_0=1-\Wc/W$ due to Coulomb interactions
where \Wc~is the minimum kinetic energy required for
collision (for protons $\Wc\approx3.5$~MeV). 
If the probability for survival at length $\Delta{l}$
equals $\Delta\Pc=1-q$, then for length $l=n\cdot\Delta l$
and small $q\ll1$ one obtains 
$\Pc(l)=(\Delta{\Pc})^n=(1-q)^n=\prod\big[1-q_0(1-\Wc/W(l))\big]$,
or
\begin{equation}
\dydx{\ln P_\mathrm{col}}{x}=-1+\frac{y_\mathrm{c}}{y}
\label{eq:Pcol}
\end{equation}
where $\midx{y}{c}=\Wc/(m_0c_0^2)$.


Let the probability for survival of unstable particles 
during the interval $\Delta{t}$ be 
$\Delta\Pd\approx 1-\Delta{t}/\tau_0$, where $\tau_0$
is the mean life time at rest. 
For relativistic particle of kinetic energy
$\Wk=(\gamma-1)m_0c_0^2$ we find
\begin{equation}
\dydx{\ln \Pd}{x}=-\frac{\lambda}{c\,\tau_0}\cdot\frac{1}{\sqrt{y^2+2y}}
\label{eq:Pdec}
\end{equation}
The average length of the path is thus
\begin{equation}
\average{l}=\int_0^L l\, \bigg\vert\dydx{P}{l}\bigg\vert\,
\mathrm{d}l
\label{eq:lavg}
\end{equation}
where $P=\Pc$ for collision and $P=\Pd$ for decay.

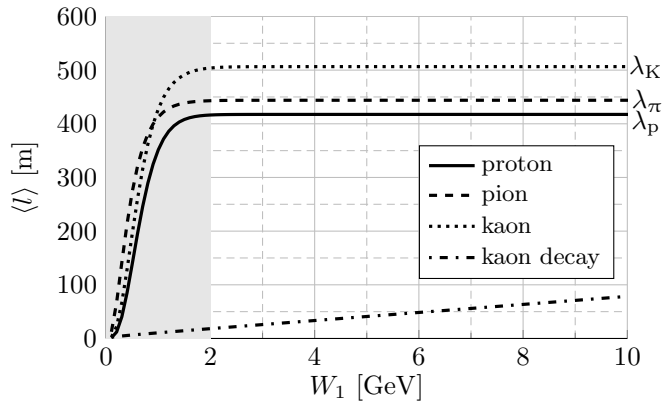
\begin{figure}[hbt]
	\centerline{\input{lavg.tikz}}
\caption{Average lengths between two successive nuclear 
	collisions for \Xp,
\Xpi, and \XK~with initial energy of 2~GeV or more is
between 415 (for protons) m and 510 m (for kaons). The shaded
part of graph shows a region of impossible outcomes: initial
energies below 2~GeV do not suffice for a particle to pass
the distance between a cloud and ground (see also
Fig.~\ref{fig:rangeqe}).}
\label{fig:avgl}
\end{figure}

The average lengths of internodes as result of a collision
and those that are a result of decay $\XK\to3\Xpi$ are
plotted in Fig.~\ref{fig:avgl}. Average
length for \XK~that decayed into three charged \Xpi s depends
strongly on the initial energy but is less likely to occur
as kaons themselves are the rarest products among the three, and
since this type of decay has a rate around 6\%.~\cite{pdg2012booklet}

\section{Discussion}

The average internode length is a mixture of averages for
all types of involved hadrons: 
$\average{l}_\Xp$ for protons, 
$\average{l}_\Xpi$ for pions, 
$\average{l}_\XK$ for kaons.
For relative rates of protons $p_\Xp$, pions $p_\Xpi$, and
kaons $p_\XK$ our prediction of the average length for a
given set of internodes is 
\begin{eqnarray*}
	\average{l}=
	p_\Xp\,\average{l}_\Xp+
	p_\Xpi\,\average{l}_\Xpi+
	p_\XK\,\average{l}_\XK
	\approx
	p_\Xp\,\lambda_\Xp+
	p_\Xpi\,\lambda_\Xpi+
	p_\XK\,\lambda_\XK
	\in[415,510]~\mathrm{m}
\end{eqnarray*}
The correctness of this prediction may suffer due to two types 
of the processes not detected when observing lightning 
geometry: first, the rare decay of kaons into three
charged pions with energy dependent average tends to lower
the average length for forkings with either two or three
prongs, and second, collisions of swap type (see reaction
type b) in Section 1.) do not produce a fork in a channel
and make the experimental average higher.

Cloud-to-ground lightnings initiated by cosmic rays are at
present indistinguishable from the ones that are not. Since
the origin of all CG lightning channels is inside a thunder-cell,
our result for internode lengths -- if confirmed -- implies
that thunder-cell is a probable source of fast elementary
particles.
We suggest that freezing of super-cooled water at rates
above 100~kg/s in a thunder-cell bounded by $0^\circ$C and
$-40^\circ$C isotherms~\cite{rakov2003lightning} 
is the most likely process
responsible for ejection of the particles. The computed
charge densities of the order $10^{-14}$~C/m imply that the
majority of the tens of coulombs of the charge transported
during lightning process originates either from the cloud or
from the ground.

\end{document}

%% file: f0.tikz
\begin{tikzpicture}[scale=0.4]
\draw[densely dotted] (-2,1) -- (-1,0.5);
\draw[solid,thick,-stealth] (-1,0.5) -- (-0.3,0.15);
\draw[solid,thick]  (-0.4,0.2) -- (0,0);
\draw[fill=black!80] (0,0) circle (0.06);
\end{tikzpicture}

%% file: f1.tikz
\begin{tikzpicture}[scale=0.4]
\draw[densely dotted] (-2,1) -- (-1,0.5);
\draw[solid,thick,-stealth] (-1,0.5) -- (-0.3,0.15);
\draw[solid,thick]  (-0.4,0.2) -- (0,0);
\draw[densely dotted] (0,0) -- (1,-0.5);
\end{tikzpicture}

%% file: f2.tikz
\begin{tikzpicture}[scale=0.4]
\draw[densely dotted] (-2,1) -- (-1,0.5);
\draw[solid,thick,-stealth] (-1,0.5) -- (-0.3,0.15);
\draw[solid,thick]  (-0.4,0.2) -- (0,0);
\draw[fill=black!80] (0,0) circle (0.06);
\draw[solid,thick] (0,0) -- (1,0.2);
\draw[densely dotted] (1,0.2) -- (2,0.4);
\draw[solid,thick] (0,0) -- (0.5,-0.6);
\draw[densely dotted] (0.5,-0.6) -- (1,-1.2);
\end{tikzpicture}

%% file: f3.tikz
\begin{tikzpicture}[scale=0.4]
\draw[densely dotted] (-2,1) -- (-1,0.5);
\draw[solid,thick,-stealth] (-1,0.5) -- (-0.3,0.15);
\draw[solid,thick]  (-0.4,0.2) -- (0,0);
\draw[fill=black!80] (0,0) circle (0.06);
\draw[solid,thick] (0,0) -- (1,0.2);
\draw[densely dotted] (1,0.2) -- (2,0.4);
\draw[solid,thick] (0,0) -- (0.5,-0.6);
\draw[densely dotted] (0.5,-0.6) -- (1,-1.2);
\draw[solid,thick] (0,0) -- (-0.1,-0.8);
\draw[densely dotted] (-0.1,-0.8) -- (-0.2,-1.6);
\end{tikzpicture}

%% file: f4.tikz
\begin{tikzpicture}[scale=0.4]
\draw[densely dotted] (-2,1) -- (-1,0.5);
\draw[solid,thick,-stealth] (-1,0.5) -- (-0.3,0.15);
\draw[solid,thick]  (-0.4,0.2) -- (0,0);
\draw[fill=black!80] (0,0) circle (0.06);
\draw[solid,thick] (0,0) -- (1,0.2);
\draw[densely dotted] (1,0.2) -- (2,0.4);
\draw[solid,thick] (0,0) -- (0.5,-0.6);
\draw[densely dotted] (0.5,-0.6) -- (1,-1.2);
\draw[solid,thick] (0,0) -- (-0.1,-0.8);
\draw[densely dotted] (-0.1,-0.8) -- (-0.2,-1.6);
\draw[solid,thick] (0,0) -- (0.8,-0.2);
\draw[densely dotted] (0.8,-0.2) -- (1.6,-0.4);
\end{tikzpicture}

%% file: range.tikz
\begin{tikzpicture}[scale=0.8]
\pgfplotsset{major grid style={dashed}} 
\pgfplotsset{minor grid style={densely dotted}}

\begin{axis}[
	legend style={at={(0.68,0.36)},
		anchor=north west,draw=black,fill=white,align=left},
	legend cell align=left,
	grid=both,
	axis x line=bottom,
	axis y line=left,
	x axis line style={-},
	y axis line style={-},
	minor x tick num=1,
	minor y tick num=1,
	xticklabel pos=left,
	xtick align=inside,
	yticklabel pos=right,
	ytick align=inside,
	x tick label style={},
	ylabel={$L~[\mathrm{km}]$},
	xlabel={$W_1~[\mathrm{GeV}]$},
	domain=0.0:10.0,
	xmin=0.0,
	xmax=10.0,
	ymin=0,
	ymax=20,
	]
	\addplot[black,solid,very thick] table[x index=1,y expr=\thisrowno{9}*1e-3] {xavg_proton+1.dat};
	\addplot[black,dashed,very thick] table[x index=1,y expr=\thisrowno{9}*1e-3] {xavg_pion+1.dat};
	\addplot[black,dotted,very thick] table[x index=1,y expr=\thisrowno{9}*1e-3] {xavg_kaon+1.dat};
	\addplot[black,dash pattern=on 1pt off 1pt on 3pt off 1pt,very thick] table[x index=1,y expr=\thisrowno{9}*1e-3] {xavg_muon+1.dat};
	\addlegendentry{{\small proton}};
	\addlegendentry{{\small pion}};
	\addlegendentry{{\small kaon}};
	\addlegendentry{{\small muon}};
\end{axis}

\end{tikzpicture}

%% file: qe.tikz
\begin{tikzpicture}[scale=0.8]
\pgfplotsset{major grid style={dashed}} 
\pgfplotsset{minor grid style={densely dotted}}

\begin{axis}[
	legend style={at={(0.60,0.90)},
		anchor=north west,draw=black,fill=white,align=left},
	legend cell align=left,
	grid=both,
	axis x line=bottom,
	axis y line=left,
	x axis line style={-},
	y axis line style={-},
	minor x tick num=1,
	minor y tick num=1,
	xticklabel pos=left,
	xtick align=inside,
	yticklabel pos=right,
	ytick align=inside,
	x tick label style={},
	xlabel={$l~[\mathrm{km}]$},
	ylabel={$q_\mathrm{e}~[10^{-14}~\mathrm{C/m}]$},
	domain=-20.0:0.0,
	xmin=-20.0,
	xmax=0.0,
	ymin=0,
	ymax=2,
	]
	\addplot[black,solid,very thick] table[x index=0,y index=3] {particles_linvWdWdl.dat};
	\addplot[black,dashed,very thick] table[x index=4,y index=7] {particles_linvWdWdl.dat};
	\addplot[black,dotted,very thick] table[x index=8,y index=11] {particles_linvWdWdl.dat};
	\addplot[black,dash pattern=on 1pt off 1pt on 3pt off 1pt,,very thick] table[x index=12,y index=15] {particles_linvWdWdl.dat};
	\addlegendentry{{\small proton}};
	\addlegendentry{{\small pion}};
	\addlegendentry{{\small kaon}};
	\addlegendentry{{\small muon}};
\end{axis}

\end{tikzpicture}

%% file: lavg.tikz
\begin{tikzpicture}[scale=1.0]
\pgfdeclarelayer{background}
\pgfsetlayers{background,main}
\pgfplotsset{major grid style={solid}} 
\pgfplotsset{minor grid style={densely dashed}}

\begin{axis}[
	legend style={at={(0.60,0.80)},
		anchor=north west,draw=black,fill=white,align=left},
	legend cell align=left,
	grid=both,
	yscale=0.75,
	axis x line=bottom,
	axis y line=left,
	x axis line style={-},
	y axis line style={-},
	minor x tick num=1,
	minor y tick num=1,
	xticklabel pos=left,
	xtick align=inside,
	yticklabel pos=right,
	ytick align=inside,
	x tick label style={},
	xlabel={$W_1~[\mathrm{GeV}]$},
	ylabel={$\average{l}~[\mathrm{m}]$},
	domain=0.0:10.0,
    	ytick={0,100,200,300,400,500,600},
    	yticklabels={0,100,200,300,400,500,600},
	xmin=0.0,
	xmax=10.0,
	ymin=0,
	ymax=600,
	]
	\begin{pgfonlayer}{background}
		\fill[gray!20] (0,0) rectangle (2,600);
	\end{pgfonlayer}
	\addplot[black,solid,very thick] table[x index=1,y index=6] {xavg_proton+1.dat};
	\addplot[black,dashed,very thick] table[x index=1,y index=6] {xavg_pion+1.dat};
	\addplot[black,dotted,very thick] table[x index=1,y index=6] {xavg_kaon+1.dat};
	\addplot[black,dash pattern=on 1pt off 3pt on 3pt off 3pt,very thick] table[x index=1,y index=5] {xavg_kaon+1.dat};
	\addlegendentry{{\small proton}};
	\addlegendentry{{\small pion}};
	\addlegendentry{{\small kaon}};
	\addlegendentry{{\small kaon decay}};
\end{axis}
	\node[] at (7.1,2.85) {$\lambda_\Xp$};
	\node[] at (7.12,3.15) {$\lambda_\Xpi$};
	\node[] at (7.1,3.6) {$\lambda_\XK$};
\end{tikzpicture}